\documentclass[%
 reprint,
 amsmath,amssymb,
 aps,
]{revtex4-2}

\usepackage{amsmath}
\usepackage{graphicx}
\usepackage{dcolumn}
\usepackage{bm}
\usepackage{hyperref}
\usepackage[dvipsnames]{xcolor}
\usepackage{multirow}
\usepackage{subfig}

\newcommand{\lya}{Lyman-$\alpha~$}
\newcommand{\TRUNC}{{\tt TRUNCATED}}
\newcommand{\DPL}{{\tt DPL}}
\newcommand{\FLOOR}{{\tt FLOOR}}

\newcommand{\RefReport}[1]{{\color{black} #1}}

\begin{document}

\preprint{APS/123-QED}

\title{Imprints of fermionic and bosonic mixed dark matter on the 21-cm signal at cosmic dawn.}

\author{Sambit K. Giri}
 \email{sambitkumar.giri@uzh.ch}
\author{Aurel Schneider}%
 \email{aurel.schneider@uzh.ch}
\affiliation{%
Center for Cosmology and Theoretical Astrophysics, \\
Institute for Computational Science, University of Zurich, Switzerland.
}%




\date{\today}

\begin{abstract}
The 21-cm signal from the epoch of cosmic dawn prior to reionization consists of a promising observable to gain new insights into the dark matter (DM) sector. In this paper, we investigate its potential to constrain mixed (cold + noncold) dark matter scenarios that are characterized by the noncold DM fraction ($f_{\rm nCDM}$) and particle mass ($m_{\rm nCDM}$). As noncold DM species, we investigate both a fermionic (sterile neutrino) and a bosonic (ultralight axion) particle. We show how these scenarios affect the global signal and the power spectrum using a halo-model implementation of the 21-cm signal at cosmic dawn. Next to this study, we perform an inference-based forecast study based on realistic mock power spectra from the Square Kilometre Array (SKA) telescope. Assuming inefficient, yet non zero star formation in minihaloes (i.e. haloes with mass below $10^8$ $M_{\odot}$), we obtain stringent constraints on both $m_{\rm nCDM}$ and $f_{\rm nCDM}$ that go well beyond current limits. Regarding the special case of $f_{\rm nCDM}\sim 1$, for example, we find a constraint of $m_{\rm nCDM}>15$ keV (thermal mass) for fermionic DM and $m_{\rm nCDM}>2\times10^{-20}$ eV for bosonic DM. For the opposite case of dominating cold DM, we find that at most 1\% of the total DM abundance can be made of a hot fermionic or bosonic relic. All constraints are provided at the 95\% confidence level. 

\end{abstract}

\maketitle


\section{\label{sec:intro}Introduction}

The cosmic dawn refers to the time before the epoch of reionization when the first stars formed and the radiation from these stars percolated the Universe. This stellar radiation induced spin flips in the neutral hydrogen atoms of the primordial gas, leading to an absorption signal in the cosmic background radiation at radio frequencies. 
Upcoming telescopes, such as the Square Kilometre Array \cite[SKA,][]{dewdney2009square, Koopmans2015TheArray} will observe this signal, thereby revealing information about the formation of first sources and the distribution of the primordial gas at redshifts between about 10 and 25 \cite{Pritchard200721-cmReionization}.

The 21-cm signal at cosmic dawn is a promising probe for studying alternative dark matter (DM) scenarios. It is particularly sensitive to DM decays or annihilation \citep{Furlanetto:2006wp,Liu:2018uzy,List:2020rrj}, interactions between DM and standard model particles \citep{Barkana2018PossibleStars,munoz2018small}, or DM models yielding a suppression of the small-scale matter power spectrum \citep{Sitwell2014TheSignal,Schneider2018ConstrainingSignal,Lidz:2018fqo,nebrin2019fuzzy,Boyarsky:2019fgp,jones2021fuzzy}. While decay and annihilation scenarios provide additional sources of radiation that may affect the 21-cm signal, interacting DM models can lead to additional cooling channels, acting on the temperature and the distribution of the gas. Models with suppressed power spectra, on the other hand, are characterized by a smaller number of sources and by a reduced clustering amplitude of the primordial gas.

In this paper, we investigate mixed dark matter scenarios featuring both a perfectly cold and a warm or hot subcomponent. Depending on the composition of the fluid (which is parametrized by the abundance ratio of hot to total dark matter) this includes all cases from the standard warm DM scenario to a $\Lambda$CDM model with an additional, hot relic particle. Note that mixed dark matter models have been studied extensively in the literature primarily in the context of the Lyman-$\alpha$ forest \citep{boyarsky2009lyman,Baur:2017stq}, Milky Way satellites \cite{Anderhalden:2012jc,Marsh:2013ywa,schneider2015structure}, strong lensing systems \citep{Kamada:2016vsc}, or the weak lensing shear signal \citep{Schneider:2019xpf, parimbelli2021mixed}. Recently, it has also been applied to the global 21-cm signal at cosmic dawn \cite{Schneider2018ConstrainingSignal,chatterjee2019ruling} in context of the claimed first detection from the Experiment to Detect the Global EoR Signal \cite[EDGES,][]{EDGES2018}. Since the interpretation of the signal from the EDGES is highly disputed \citep[see e.g.][]{Hills2018ConcernsData,singh2019redshifted}, we do not discuss it further in this paper.

From a particle physics perspective, mixed dark matter scenarios can emerge from a variety of different contexts. One popular option is an additional, right-handed (sterile) neutrino species with a particle mass above the electron volt (eV) range \citep{Drewes:2016upu}. Such a fermionic particle can be produced in the early Universe via freeze-in \citep{Dodelson:1993je, Shi:1998km} or decay production \citep{Petraki:2007gq,Merle:2013wta}, acting as a hot/warm DM subcomponent similar to the left-handed neutrinos. Another option is an ultralight axionlike particle \citep{Hu:2000ke,Hui2021WaveDM,schwabe2020simulating}. Such bosonic particles could be abundantly present in the early Universe, some of them featuring suppressed structure formation due to a de Broglie wavelength of astrophysical relevance. Finally, mixed dark matter scenarios with suppressed power spectra can also arise if some of the DM interacts with another dark relic or with particles from the standard model in the early Universe. Next to the characteristic suppression in power, such models could also feature remnants of acoustic oscillations \citep{Cyr-Racine:2013fsa,Bohr:2020yoe,Schaeffer:2021qwm}.

In this paper we focus on mixed scenarios featuring either bosonic (axionlike) or fermionic (neutrinolike) dark matter. In particular, we investigate the imprint of these models on the spatial fluctuations of the 21-cm signal at cosmic dawn. Although current interferometry-based radio telescopes, such as the Low Frequency Array \citep[LOFAR,][]{vanHaarlem2013LOFAR:ARray}, the Murchison
Widefield Array \citep[MWA,][]{wayth2018phase}, or the Hydrogen Epoch of Reionization Array \cite[HERA,][]{deboer2017hydrogen} have only provided upper limits on the 21-cm power spectrum at cosmic dawn \cite[see e.g.][]{gehlot2020aartfaac,Yoshiura2021new}, the low frequency component of the upcoming SKA (SKA-Low) is expected to detect the signal at high significance. In this paper, we therefore specifically focus on a SKA-Low setup, forecasting the power of the SKA telescope to detect or constrain fermionic or bosonic mixed dark matter.

 \begin{figure*} 
 \centering
  \includegraphics[width=0.9\textwidth]{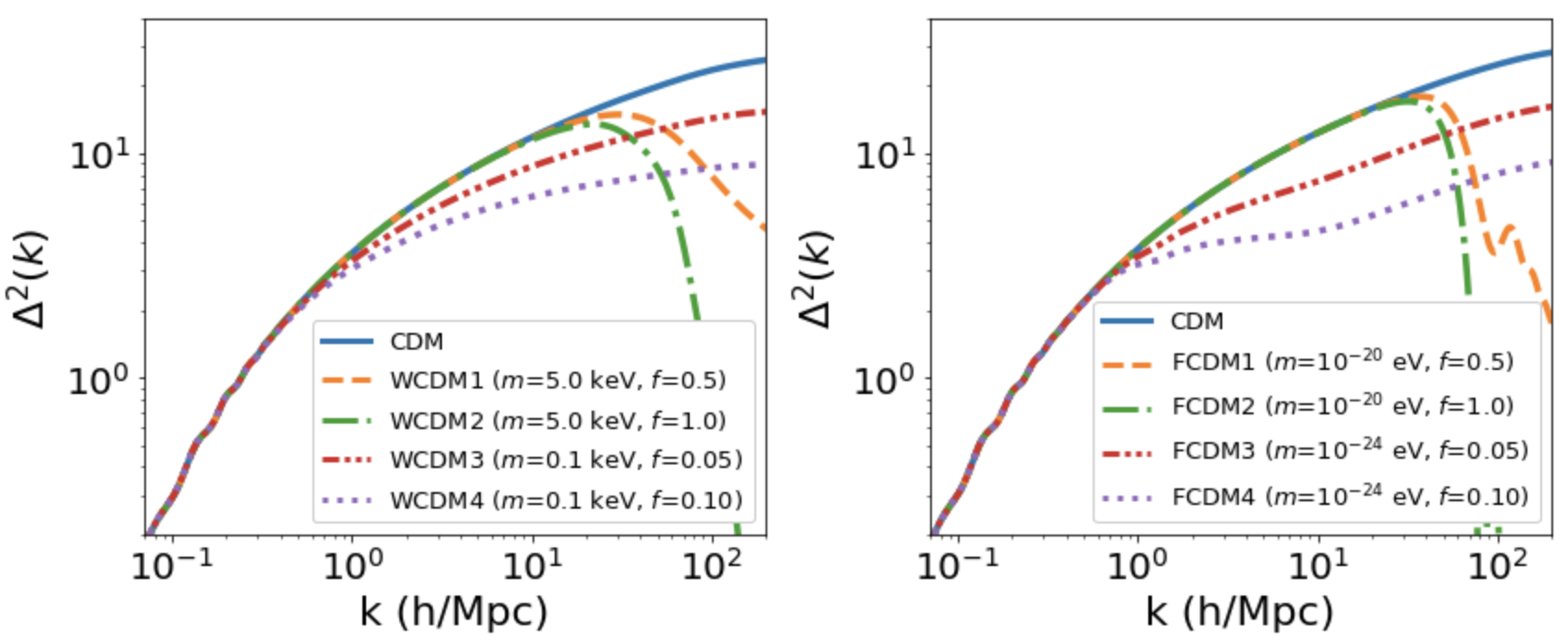}
 \caption{The linear dimensionless matter power spectra for various models at $z=0$. In the left- and right-hand panels, we focus on fermionic and bosonic mixed DM models, respectively. The green lines show models with $f_{\rm nCDM}=1$, i.e. pure warm DM with $m_\mathrm{WDM}= 5$ keV and fuzzy DM with $m_\mathrm{FDM}= 10^{-20}$ eV. The orange lines correspond to the same mixed DM models with $f_{\rm nCDM}=0.5$. In red and purple, we show models with small fractions ($f_{\rm nCDM}=0.05$, $f_{\rm nCDM}=0.1$) and particle masses ($m_\mathrm{WDM}= 0.1$ keV, $m_\mathrm{FDM}= 10^{-24}$ eV). The CDM case is shown as a blue solid line for comparison. Note that bosonic DM models are characterized by steeper, more abrupt downturns compared to fermionic DM models.} \label{fig:matter_ps_DMmodels}
 \end{figure*}

We begin by describing our parameterization for the DM and source scenarios in Secs.~\ref{sec:dm_model} and~\ref{sec:source_model}, respectively. 
Section~\ref{sec:21cm_signal} is dedicated to the resulting 21-cm global signal and power spectrum for mixed dark matter scenarios. In Sec.~\ref{sec:forecast}, we present a forecast study for the SKA telescope. Finally, we summarize our findings in Sec.~\ref{sec:summary}.

\section{\label{sec:dm_model} Mixed dark matter scenarios}
 
In a two-component mixed dark matter (mixed DM) framework, the DM sector consist of a  perfectly cold plus a noncold component. By noncold DM we either mean a hot or warm fermionic DM fluid with large free-streaming velocities or a very light bosonic particle with wave properties affecting astrophysical scales. The fraction of noncold DM is given by
\begin{eqnarray}
f_{n\mathrm{CDM}} = \frac{\Omega_{\mathrm{nCDM}}
}{\Omega_{\mathrm{nCDM}}+\Omega_{\mathrm{CDM}}} \ ,
\end{eqnarray}
where $\Omega_{\mathrm{nCDM}}$ and $\Omega_{\mathrm{CDM}}$ describe the cosmological abundances of the two fluids. Note that $f_{\rm nCDM}=1$ corresponds to the fully warm or fuzzy DM scenarios, while $f_{\rm nCDM}\ll 1$ denotes the case of a $\Lambda$CDM model with a very subdominant, additional noncold species.

Hereafter, the fermionic mixed DM scenario is denoted as WCDM, which stands for warm and cold dark matter. Next to the warm-to-total fraction ($f_{\rm WCDM}$), the model is characterized by the mass of the fermionic relic ($m_{\rm WDM}$). Following the historical convention, we use the mass definition of a thermally produced particle. This mass is different from the true particle mass in the case of nonthermal production in the early Universe. Note, however, that there is a direct mapping between the different mass definitions, as shown, e.g., by Ref.~\cite{viel2005constraining}.

The fermionic mixed DM scenario is best motivated by the presence of an additional sterile neutrino species. Sterile neutrinos could have been produced in the early Universe via mixing with the active neutrino sector. This process, known as Dodelson-Widrow production \cite{Dodelson:1993je}, is ruled out for scenarios with $f_{\rm WCDM}=1$ \cite{Seljak:2006qw,viel2005constraining} but could be present if sterile neutrinos only make up a fraction of the total DM sector. Other mechanisms discussed in the literature are the resonant \cite[or Shi-Fuller,][]{Shi:1998km} and the decay production. Note that these scenarios may still be viable even for the case of $f_{\rm nCDM}=1$, although  large fractions of their parameter space have recently been ruled out \cite[e.g.][]{Schneider:2016uqi,Perez:2016tcq}.

The bosonic mixed DM scenario is abbreviated as FCDM, standing for fuzzy and cold dark matter. The model is characterized by the fuzzy-to-total DM fraction $f_{\rm FCDM}$ and the bosonic particle mass $m_{\rm FDM}$. The bosonic particle fluid forms a condensate in the early Universe, leading to a suppression of the particle free-streaming process \cite{Hu:2000ke,Marsh:2013ywa}. As a consequence, models with $f_{\rm FCDM}=1$ and tiny particle masses of order $m_{\rm FDM}\sim10^{-21}$ eV evade constraints from small-scale structure formation. Note that bosonic particles of this mass range are characterized by their wavelike nature that affects astrophysical scales and leads to the typical small-scale power suppression of mixed DM models.


In Fig. \ref{fig:matter_ps_DMmodels}, we show the dimensionless matter power spectra of fermionic (left panel) and bosonic (right panel) DM benchmark models at $z=0$. The full WDM and FDM cases with $f_{\rm nCDM}=1$, and with masses of $m_{\rm WDM}=5$ keV and $m_{\rm FDM}=10^{-20}$ eV, respectively, are shown as green dashed-dotted lines. These models exhibit a strong cutoff at large  $k$ values compared to the standard CDM model shown as a blue solid line. As expected, the cutoff is significantly steeper for fuzzy DM compared to the warm DM model \cite{Marsh:2013ywa,Murgia2017Non-coldApproach}.

The orange lines in Fig.~\ref{fig:matter_ps_DMmodels} show the power spectra of WCDM (left) and FCDM models (right) with $f_{\rm nCDM}=0.5$. The particle masses are $m_{\rm WDM}=5$ keV and $m_{\rm FDM}= 10^{-20}$ eV, i.e., the same as for the full WDM and FDM cases shown in green. Due to the reduced noncold fraction, the visible power cutoff is significantly shallower with some remaining power down to small scales (high $k$ values).

Next to models featuring fairly strong downturns in the power spectrum, we also show examples which are only slightly damped toward higher $k$ values. These models are characterized by small particle masses ($m_{\rm WDM}=0.1$ keV and $m_{\rm FDM}=10^{-24}$ eV) and small fractions ($f_{\rm nCDM}=0.05$ and $f_{\rm nCDM}=0.10$), as shown by the red dashed-dotted-dotted and purple dotted lines in Fig.~\ref{fig:matter_ps_DMmodels}. These models are qualitatively different from the ones with higher fractions because they affect the source abundance at all scales and yield modified power spectra at $k$ modes that are directly observable with future radio interferometric surveys.

Note that throughout this paper, we use the Boltzmann solver \texttt{class} \cite{blas2011classII,lesgourgues2011classIV} to calculate the WCDM power spectra. For the FCDM power spectra, we rely on \texttt{axionCAMB} \cite{hlovzek2018using}, which is a modified version of \texttt{CAMB} \cite{lewis2011camb} that includes the option of an additional axionlike DM component.

\section{\label{sec:source_model} Source model}

One of the main reasons why the 21-cm signal can be used to distinguish between cold and mixed DM models, is the change of source distributions between the two models. The assumed properties of sources are therefore a crucial model component, which we discuss in the present section.

\subsection{Halo mass function, bias, and growth}

We assume all photon sources to form within DM haloes. The halo number density is described by the halo mass function obtained from the extended Press-Schechter (EPS) approach,
\begin{eqnarray}
\frac{\mathrm{d} n}{\mathrm{d} \mathrm{ln} M} =-\frac{\bar{\rho}}{M}\nu f(\nu) \frac{\mathrm{d} \mathrm{ln}\sigma}{\mathrm{d} \mathrm{ln} M} \ ,
\end{eqnarray}
where $\nu = (\delta^2_c/\sigma)^2$ with $\delta_c = 1.686$ and $\bar{\rho}$ is the mean baryon density. The halo mass $M$ is given by $M=(4\pi/3)\bar{\rho}(cR)^3$, where $c$ is a free parameter that has to be fixed to simulations. We assume the first-crossing distribution $f(\nu)$ to have the following form \cite{sheth1999large},
\begin{eqnarray}
f(\nu) = A\sqrt{\frac{2q\nu}{\pi}}(1+\nu^{-p})e^{-q\nu/2} \ ,
\end{eqnarray}
with $A=0.3222$, $p=0.3$, and $q=1$. The variance is given by
\begin{eqnarray}
\sigma^2(R, z) = \int \frac{\mathrm{d}k^3}{(2\pi)^3} P_\mathrm{lin}(k)\mathcal{W}(k|R) \ ,
\end{eqnarray}
where we assume the window function to be described by the smooth-$k$ filter \cite{leo2018new}
\begin{eqnarray}
\mathcal{W}(k|R) = \frac{1}{1+(kR)^\beta} \ .
\end{eqnarray}
Following Ref.~\cite{parimbelli2021mixed}, we assume $c=3.3$ and $\beta=4.8$, as these values provide the best match to numerical simulations of a series of mixed DM models.

\begin{figure*}[t] 
 \centering
~~~~~~~~~\textbf{\large WCDM}
\includegraphics[width=0.98\textwidth]{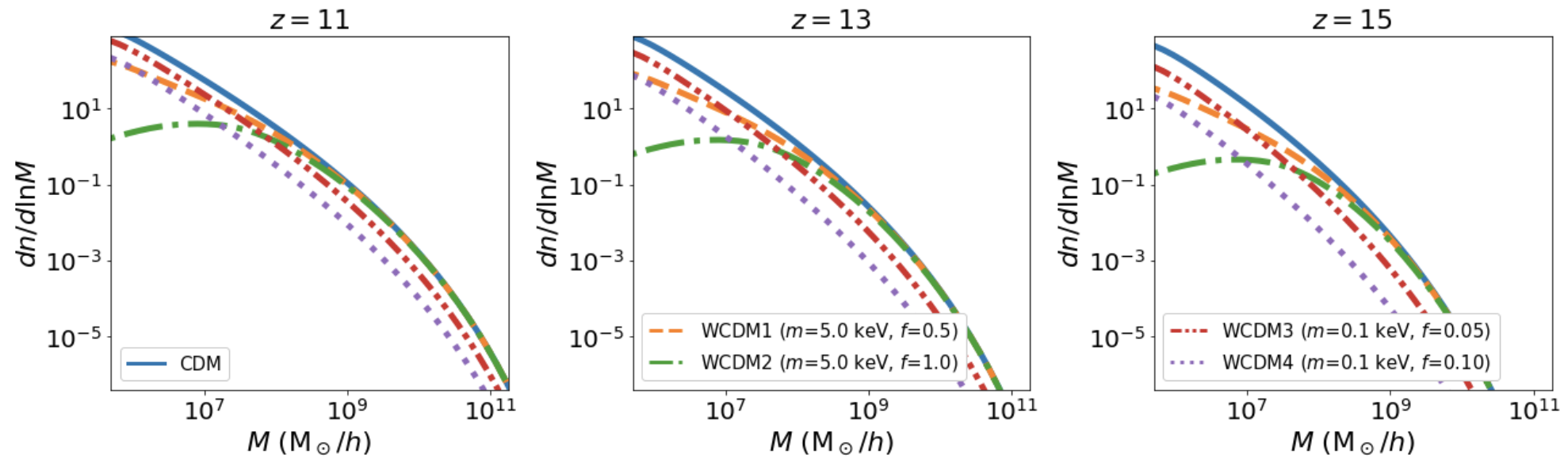}\\ \vspace{0.3cm} 
~~~~~~~~~\textbf{\large FCDM}
\includegraphics[width=0.98\textwidth]{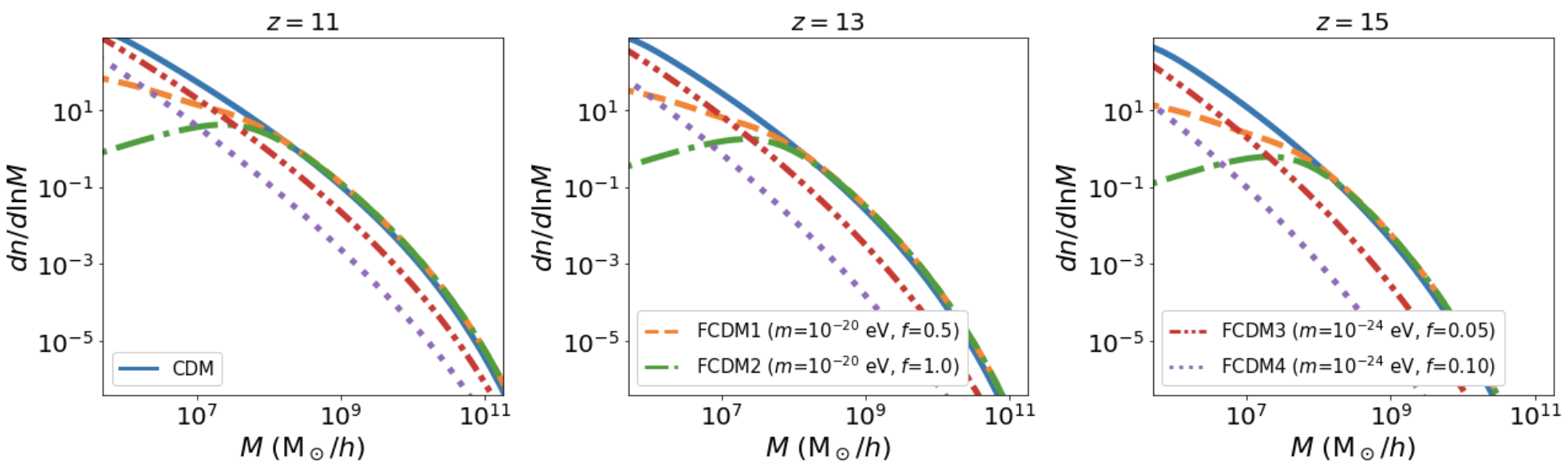}
 \caption{The halo mass function of different fermionic (top panels) and bosonic (bottom panels) mixed DM scenarios at redshift $z =$ 11 (left panels), 13 (middle panels), and 15 (right panels). The shown models are the same ones shown in Fig.~\ref{fig:matter_ps_DMmodels}. 
 }
 	\label{fig:hmf_cwdm_cfdm}
 \end{figure*}

In Fig.~\ref{fig:hmf_cwdm_cfdm}, we plot the halo mass functions of the benchmark WCDM (top panels) and FWDM (bottom panels) models shown in Fig.~\ref{fig:matter_ps_DMmodels}. The different columns correspond to the three different redshifts $z=11$ (left), $13$ (middle), and $15$ (right). As expected, the full WDM ($m_{\rm WDM} =5$ keV) and FDM ($m_{\rm FDM}=10^{-20}$ eV) models exhibit a strongly suppressed halo abundance below $M\sim 10^9$ $M_{\odot}$/h compared to the CDM model. The corresponding mixed DM models (with $f_{\rm nCDM}=0.5$) also show suppressed halo numbers, but the mass functions keep rising toward small halo masses. 
The two models with low fraction ($f_{\rm nCDM}=0.05,\,0.1$) and small particle mass ($m_{\rm WDM}=0.1$ keV, $m_{\rm FDM}=10^{-24}$ eV) are characterized by a low halo number density over the full mass range. Compared to the CDM model, the halo mass functions are shifted downwards while retaining a similar overall shape. \RefReport{Even though these models do not reproduce the CDM halo number density at any mass scale, they remain in agreement with low redshift observations \cite[e.g.][]{schneider2015structure,boyarsky2009lyman,baur2017constraints,davoudiasl2021ultralight}.}

Using the peak-background split model of Ref.~\cite{sheth1999large}, the halo bias can be described within the EPS formalism. The halo bias is given as
\begin{eqnarray}
b(M) = 1 + \frac{q\nu - 1}{\delta_c(z)} + \frac{2p}{\delta_c (z)[1+(q\nu)^{p}]} \ ,
\end{eqnarray}
where $q$ and $p$ have been defined above, and $\delta_c(z)=\delta_c/D(z)$, with $D(z)$ being the growth function \cite[see e.g. Ref][]{Cooray2002HaloStructure}. \RefReport{The halo bias is one of the key ingredients required to calculate the power spectrum in the halo model \cite[see eq.~35 in Ref.~][]{Schneider2021HaloDawn} and changes for different nCDM models \cite[e.g.][]{schneider2012non}.}

To describe the growth of halos, we assume a halo mass accretion model based on the EPS method. In this method, the accretion mass $M_\mathrm{ac}$ is given by
\begin{eqnarray}
\frac{\mathrm{d} M_\mathrm{ac}}{\mathrm{d}z} = - \sqrt{\frac{2}{\pi}} \frac{M_\mathrm{ac}}{\sqrt{\sigma^2(QM_\mathrm{ac},0)-\sigma^2(M_\mathrm{ac},0)}} \frac{\mathrm{d}\delta_c(z)}{\mathrm{d}z} \ .
\end{eqnarray}
The above differential equation can be solved assuming $M_\mathrm{ac}(M, z_f) = M$, where $z_f$ designates the final redshift of interest. $Q$ is a free parameter. We choose $Q=0.6$ as it agrees with the simulations in Ref. \cite{behroozi2020universe}.
See Ref.~\cite{Schneider2021HaloDawn} for a more detailed description of the mass accretion model.

\subsection{Stellar-to-halo mass relation}
For now, we have established a model for the halo growth and distribution. What remains to be modeled is the connection of these haloes to photon emitting sources. Following Ref.~ \cite{Schneider2021HaloDawn}, we assume the stellar-to-halo mass relation,
\begin{eqnarray}\label{shmr}
f_\star (M) = \frac{2(\Omega_b/\Omega_m)f_{\star,0}}{\left(\frac{M}{M_p}\right)^{\gamma_1} + \left(\frac{M}{M_p}\right)^{\gamma_2}} \times S(M) ,
\label{eq:fstar_M}
\end{eqnarray}
which consists of a double power law multiplied by a small-scale term
\begin{eqnarray}\label{shmrsmall}
S(M) = \left[1+\left(\frac{M_t}{M}\right)^{\gamma_3}\right]^{\gamma_4}
\label{eq:S_M}
\end{eqnarray}
that either leads to a boost or the a further suppression of the stellar-to-halo mass relation below $M_t$.

For the double-power law of Eq.~(\ref{shmr}), we assume $M_p=2.5\times10^{11}$ $M_{\odot}$/h, $\gamma_1=0.49$, $\gamma_2=-0.61$ \cite{Mirocha2017TheFunction}. These numbers guarantee the recovery of the observed high-redshift luminosity functions of Refs.~\cite[][]{bouwens2015uv,finkelstein2015evolution} assuming a $\Lambda$CDM universe.

The small-scale term of Eq.~(\ref{shmrsmall}) provides a parametrization of the uncertainties related to the stellar-to-halo mass relation at small mass scales. 
\RefReport{For large masses ($M_t\gtrsim 10^9$ M$_{\odot}/h$), we have constraints from high redshift ($6\lesssim z \lesssim 11$) observations of luminosity functions abundance matched to the CDM haloes \cite{Mirocha2017TheFunction}. However, we currently have no constraints on the population of galaxies residing in the small mass haloes below $\sim 10^9$ M$_{\odot}/h$.}

Accounting for the constraints discussed above, we assume a stellar-to-halo mass with free parameters $f_{\star,0}$ $M_t$, $\gamma_3$, and $\gamma_4$. We furthermore restrict a hard limit of $f_{\star}(M)\leq \Omega_b/\Omega_m$ imposed by the simple constraint that the galaxy is a part of the total halo mass. Finally, we assume the stellar-to-halo mass relation to drop to zero below $M_{\rm min}=5\times 10^5 M_\odot$/h, which is motivated by the molecular cooling limit \cite{tegmark1997small}. \RefReport{The molecular cooling limit corresponds to a virial temperature of 500 K \cite{tegmark1997small}, which translates to a halo mass of $\sim 7\times 10^5 M_\odot$/h and $\sim 3\times 10^5 M_\odot$/h at redshift 10 and 20 respectively, following eq.~26 in Ref.~\cite{Barkana2001InUniverse}. Note, however, that this is an approximate relation in a CDM universe and the conversion is tricky for nCDM universe \cite[e.g.][]{safarzadeh2018limit,Lidz:2018fqo,Schneider2018ConstrainingSignal,lopez2019dark}. Therefore for simplicity, we use a redshift independent $M_\mathrm{min}$ in this work.}

\begin{figure}[t] 
 \centering
  \includegraphics[width=0.45\textwidth]{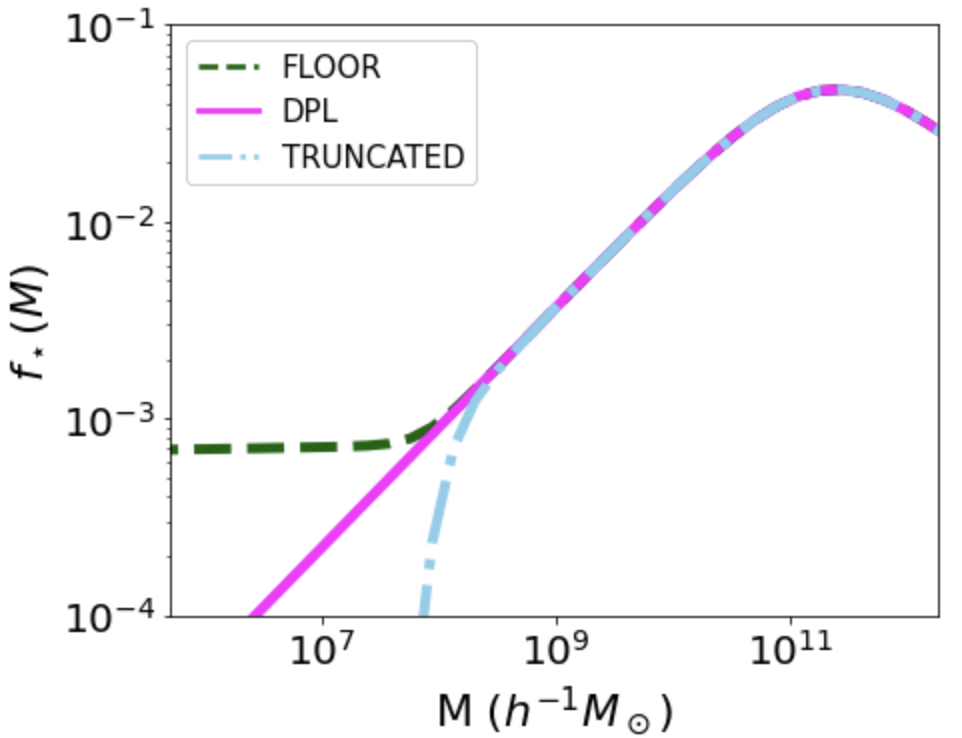}
 \caption{The fraction of halo mass converted into photon emitting stars plotted against the halo masses. We show the three models explored in this study (see text).}
 	\label{fig:source_models}
\end{figure}

In Fig.~\ref{fig:source_models}, we provide three examples of possible stellar-to-halo mass relations using the parametrization described above. The magenta solid line shows the case of 
$\gamma_4=0$, resulting in a double power law shape all the way down to the molecular cooling limit (\DPL\ model). The sky blue dashed-dotted line is characterized by a strong truncation around the scale where atomic cooling becomes inefficient (\TRUNC\ model). The corresponding parameters are $M_t=7\times 10^7$ $M_{\odot}$/h, $\gamma_3=3$, and $\gamma_4=-3$ \cite[see Refs.][for a justification of the truncation mass used here]{Barkana2001InUniverse,Dixon2016TheReionization}. The dark green dashed line finally provides an example of a stellar-to-halo mass relation that is flattening toward small scales (\FLOOR\ model). Such a scenario could be realized via the efficient formation of first generation (pop III) stars. \RefReport{See Refs.~\cite{wise2014birth,xu2016galaxy} for examples of such a model. In this work, the \FLOOR\ model is defined by} the parameters $M_t=7\times 10^7$ $M_{\odot}$/h, $\gamma_3=3$, and $\gamma_4=0.2$.

The three source models \DPL, \TRUNC, and \FLOOR\ presented in Fig.~\ref{fig:source_models} are examples that stand for our currently poor knowledge of the star-formation process in minihaloes. Later in the paper, they are used as benchmark source models for our SKA mock observations.

\subsection{Spectral emission}
After having established the connection between sources and haloes, we still have to set up a parametrization for the spectral energy distribution (SED) of the source emission. For both the SED of the Lyman-$\alpha$ ($\alpha$) and x-ray ($X$) flux, we assume a power law
\begin{eqnarray}
I_s(\nu) = A_s \nu^{-\alpha_s} \ ,
\end{eqnarray}
where $s=\{ \alpha, X \}$. We fix the spectral indices to $\alpha_{\alpha}=0$ and $\alpha_{X}=1.5$, respectively. The normalization constants $A_{s}$ are set such that
\begin{eqnarray}
\int^{\nu_{2,s}}_{\nu_{1,s}} I_s (\nu) = 1 \ .
\end{eqnarray}
For the Lyman-$\alpha$ flux, the integration limits are given by $\nu_{1,\alpha}=2.467\times 10^{15}$ Hz and $\nu_{2,\alpha}=3.290\times 10^{15}$ Hz, which correspond to the Lyman-$\alpha$ and Lyman-limit frequencies. For the x-ray integration limits, we assume  $\nu_{1,X}= E_0/h_P$ and $\nu_{2,X}=2~\mathrm{keV}/h_P$, where $E_0$ is a free model parameter. See, e.g., Ref.~\cite{greig2017simultaneously} for a discussion on dependence of the 21-cm power spectrum on $E_0$.

The number emissivity of UV
photons between the Lyman-$\alpha$ and Lyman-limit range is given by
\begin{eqnarray}
\epsilon_\alpha (\nu) = \frac{N_\alpha}{m_p} I_\alpha (\nu) \ ,
\end{eqnarray}
where $N_\alpha$ is the number of photons per baryon emitted
in the range between the Lyman-$\alpha$ ($\nu_\alpha$) and the Lyman limit ($\nu_\mathrm{LL}$) frequencies. The energy emissivity of x-ray
photons is
\begin{eqnarray}
\epsilon_X(\nu) = f_X c_X \frac{I_X(\nu)}{\nu h_\mathrm{P}}
\end{eqnarray}
where $f_X$ is a free parameter of order unity, and $c_X$ is
a normalization factor constrained by observations. It is set to $c_X = 3.4 \times 10^{40}$ 
${\rm erg}~{\rm yr}~s^{-1}~M^{-1}_\odot$
based on the findings of Ref.~\cite{gilfanov2004lx}.



 \begin{figure*}[t] 
 \centering
 {\large ~~~~~~~~~~~~~~\FLOOR~~~~~~~~~~~~~~~~~~~~~~~~~~~~~~~~~~~~~\DPL~~~~~~~~~~~~~~~~~~~~~~~~~~~~~~~~~~\TRUNC}
  \includegraphics[width=1.0\textwidth]{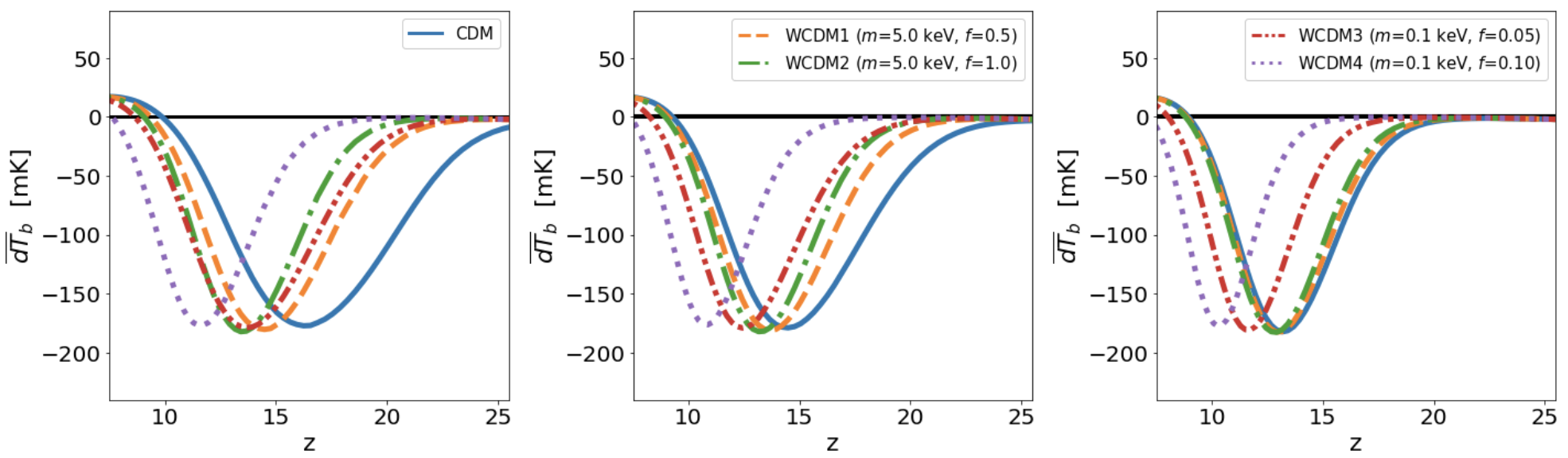}
  \includegraphics[width=1.0\textwidth]{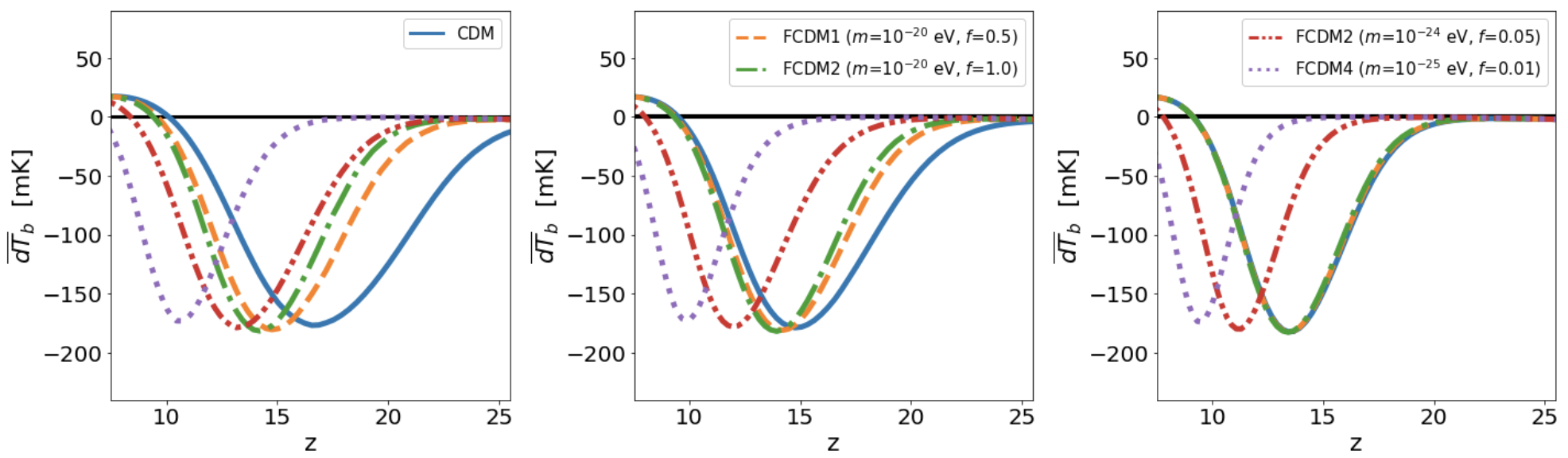}
 \caption{Sky-averaged 21-cm differential brightness temperature ($\bar{dT}_\mathrm{b}$) for CDM and various WCDM (top panels) and FCDM (bottom panels) models. We assume three different source models characterized by a flat (\FLOOR), a power law (\DPL), and a truncated stellar-to-halo mass relation at small scales (see Fig.~\ref{fig:source_models}).}
 	\label{fig:global_signal}
 \end{figure*}

\section{21-cm Signal}\label{sec:21cm_signal}
In this section, we provide a short summary of the 21-cm signal focusing on the effects induced by a mixed DM sector. We discuss both the global signal and the power spectrum with a primary focus on the latter. All calculations are performed using the model presented in Ref.~\cite{Schneider2021HaloDawn}. We only discuss some general terminology here and refer to Ref.~\cite{Schneider2021HaloDawn} for any details regarding the formalism.

The observed 21-cm signal is given by the differential brightness temperature $\delta T_b(\mathbf{x},z)$ which is a function of position $\mathbf{x}$ and redshift $z$,
\begin{eqnarray}\label{eq:dTb}
\delta T_b(\mathbf{x},z)=T_0(z)x_{\rm HI}(\mathbf{x},z)\left[1+\delta_b(\mathbf{x},z)\right]\nonumber\\
\frac{x_{\alpha}(\mathbf{x},z)}{1+x_{\alpha}(\mathbf{x},z)}\left(1-\frac{T_{\rm cmb}(z)}{T_{k}(\mathbf{x},z)}\right),
\end{eqnarray}
where $T_{\rm cmb}$ is the cosmic microwave background (CMB) temperature, $x_{\rm HI}$ the neutral hydrogen fraction, and $\delta_b$ describes the HI gas perturbation field that is assumed to follow the dark matter. The prefactor $T_0$ is given by, 
\begin{eqnarray}
T_0(z) \approx 27 \left( \frac{\Omega_b h^2}{0.023}\right) \left(\frac{0.15}{\Omega_{\rm DM} h^2} \frac{1+z}{10}\right)^{1/2} {\rm mK}\ ,
\end{eqnarray}
where $\Omega_b$, $\Omega_{\rm DM}$, and $h$ are the cosmological abundance of baryons, the total DM ($\Omega_{\rm nCDM}+\Omega_{\rm CDM}$), and the Hubble parameter. Throughout this study, we assume  $\Omega_b=0.049$, $\Omega_{\rm DM}=0.315$, and $h=0.673$ \cite{PlanckCollaboration2018PlanckParameters}.

The radiation coupling coefficient is
\begin{eqnarray}
x_\alpha = \frac{1.81\times 10^{11}}{(1+z)} S_\alpha J_\alpha \ ,
\end{eqnarray}
where $S_{\alpha}$ is a correction factor of order unity \cite[calculated as indicated in Ref.][]{Furlanetto:2006jb} and $J_\alpha$ is the redhsift and position-dependent Lyman-$\alpha$ flux radiation field induced by the first stellar light. The temperature of the gas $T_k$ is obtained via the differential equation,
\begin{align}
\frac{3}{2}\frac{\mathrm{d}T_k}{\mathrm{d}z} = \frac{T_k}{\rho} \frac{\mathrm{d}\rho}{\mathrm{d}z} - \frac{\Gamma_h}{k_B(1+z) H(z)} ,
\label{eq:dTkdz_x}
\end{align}
where $\rho$ is the total matter density field and $\Gamma_h$ is the redhsift and position-dependent heating term induced by the x-ray radiation of the first sources. Details on how to calculate $J_{\alpha}$ and $\Gamma_h$ can be found in Ref. \cite{Schneider2021HaloDawn} and references therein.

\subsection{Global signal}
The global differential brightness temperature is obtained by averaging over the spatial component of Eq.~(\ref{eq:dTb}). We refer to Ref.~\cite{Pritchard200721-cmReionization} for the details of this calculation. Here we just want to note that the mean Lyman-$\alpha$ flux radiation $J_{\alpha}$ and the mean x-ray heating term $\Gamma_h$ directly depend on the star formation rate density (SFRD)
\begin{eqnarray}
\dot{\rho}_\star (z) = \int \frac{dM}{M} \frac{dn}{d\mathrm{ln}M} f_\star(M) \dot{M}_\mathrm{ac}(M,z) \ .
\end{eqnarray}
\RefReport{The above integral depends on the halo mass function and the mass accretion rate ($\dot{M}_\mathrm{ac}$). We assume the EPS method in Ref.~\cite{Schneider2021HaloDawn} to model the mass accretion.
Both the halo mass function and the mass accretion rate depend on the nature of DM.} This means that any variation of the fraction $f_{\rm nCDM}$ and mass $m_{\rm nCDM}$ parameters, which are introduced in Sec.~\ref{sec:dm_model} will directly affect the global signal.

The high sensitivity of the global 21-cm signal to DM models featuring a suppressed perturbations has been demonstrated multiple times in the past \cite[e.g.][]{Barkana:2001gr,Sitwell2014TheSignal,Safarzadeh:2018hhg,Schneider2018ConstrainingSignal,nebrin2019fuzzy,Boyarsky:2019fgp,munoz2021ethos}. Here, we limit ourselves to qualitative discussion including the DM and source models introduced in Sec.~\ref{sec:dm_model} and \ref{sec:source_model} respectively. 


 \begin{figure*}[t] 
 \centering
  {\large ~~~~~~~~~~~~~~\FLOOR~~~~~~~~~~~~~~~~~~~~~~~~~~~~~~~~~~~~~\DPL~~~~~~~~~~~~~~~~~~~~~~~~~~~~~~~~~~\TRUNC}
  \includegraphics[width=1.0\textwidth]{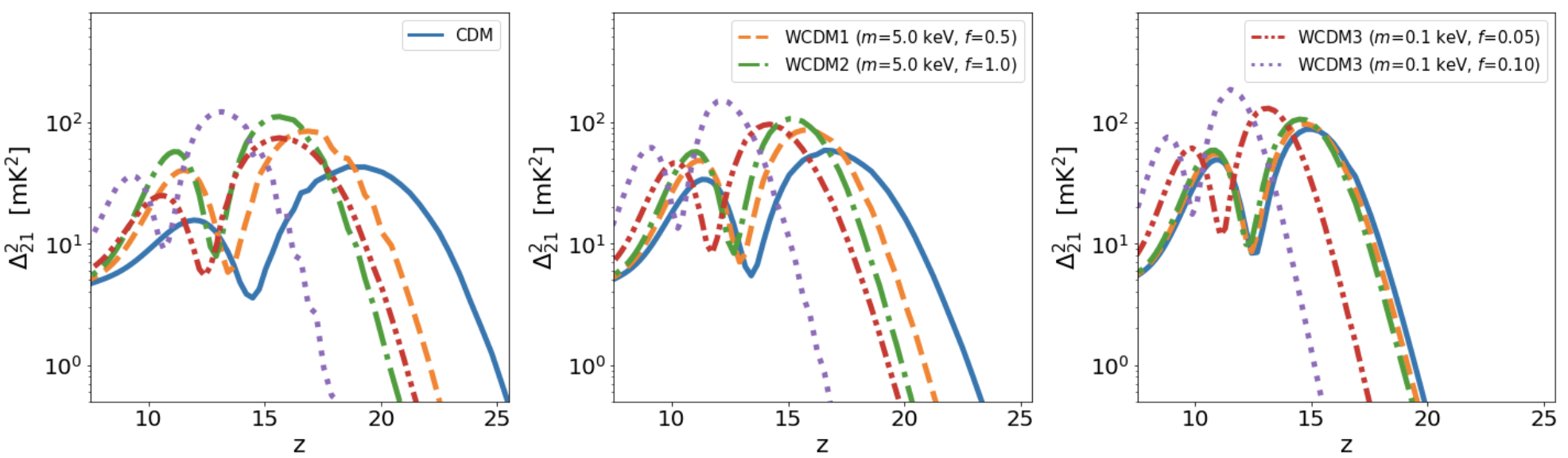}
  \includegraphics[width=1.0\textwidth]{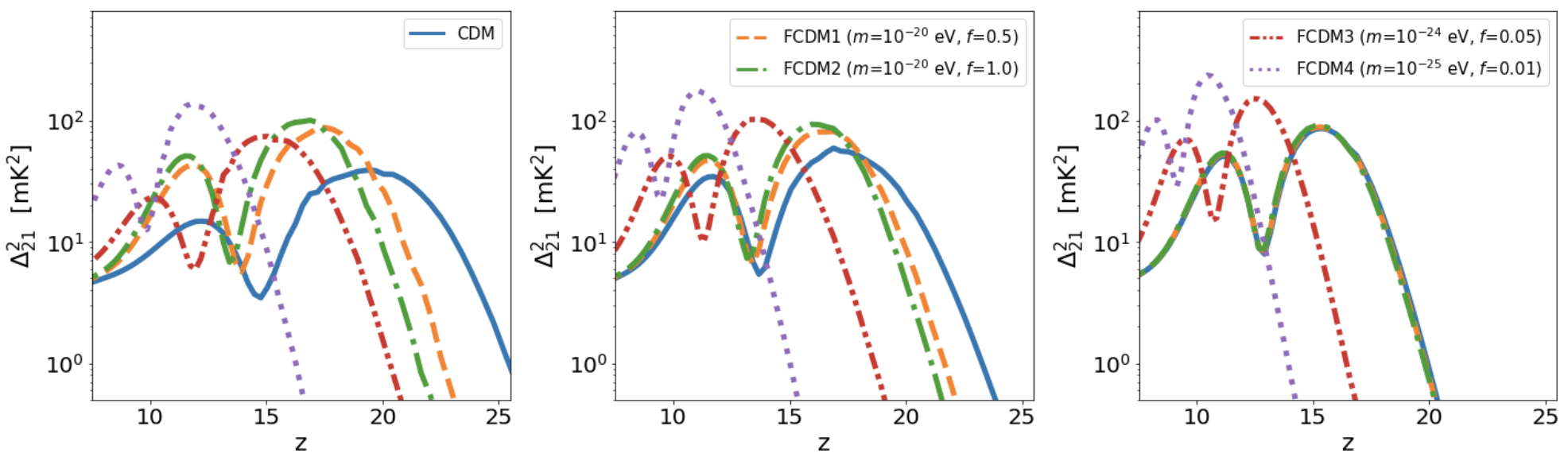}
 \caption{Evolution of the power spectra over redshift at $k=0.1~h/\mathrm{Mpc}$ for CDM and various benchmark WCDM (top panels) and FCDM models (bottom panels). We again assume the three source models: \FLOOR, \DPL, and \TRUNC\ (left to right).}
 	\label{fig:PS_vs_z}
 \end{figure*}
 
In Fig.~\ref{fig:global_signal} we plot the global differential brightness temperature for the WCDM (top panels) and FCDM (bottom panels) models assuming the \FLOOR\ (left panels), \DPL\ (central panels), and \TRUNC\ (right panels) source models. The color scheme is identical to the one used in Figs.~\ref{fig:matter_ps_DMmodels} and \ref{fig:hmf_cwdm_cfdm}.
First, we notice that the signal from the \TRUNC\ source model is shifted toward lower redshifts compared to the \DPL\ and \FLOOR\ models. This trend is a direct consequence of the reduced stellar mass in small haloes. The same is true when comparing the mixed DM models to the CDM case (shown in blue). The suppressed power spectra lead to a delay in the buildup of sources that shifts the global 21-cm signal to small redshifts. Furthermore, it is worth noticing that the \FLOOR\ source model provides the largest differences between the various DM models. This indicates that different DM models can be distinguished best from each other if the smallest haloes are populated with sources.

Finally, one should note that the DM and source models can conspire and give a similar signal. For example, the \DPL\ source model in CDM is nearly indistinguishable to the \FLOOR\ model in WCDM or FCDM scenarios, which shows that parameters related to the DM sector and the source model can be degenerate. We discuss more about these degeneracies in Sec. \ref{sec:forecast}.

\subsection{Power spectrum}
\label{sec:results_21ps}
The power spectrum of the spatially varying differential brightness temperature field [given by Eq.~(\ref{eq:dTb})] is calculated using the halo model of cosmic dawn~\cite{Schneider2021HaloDawn}. The model is build upon overlying flux profiles which include spectral redshifting and look-back light-cone effects. It accounts for the auto- and cross-spectra of all individual components (Lyman-$\alpha$ coupling, temperature, and gas density fields) and includes redshift-space distortion effects at the linear level \cite{kaiser1987clustering,Bharadwaj2004TheReionization}. The halo model of cosmic dawn has been shown to agree with other predictions based on {\tt 21cmFAST} \citep{Mesinger201121cmfast:Signal,Murray2020}.

In terms of warm, fuzzy, and mixed DM, the 21-cm power spectrum is influenced by two main effects: first, the number density and mass accretion rate of sources at the smallest mass scales and, second, the direct modifications of the matter power spectrum in the observable range. The latter is only relevant for models with small mass ($m_{\rm nCDM}$) and fraction ($f_{\rm nCDM}$), where the power spectrum is suppressed at scales of $k\lesssim 1$ h/Mpc.

Here we focus on the spherically averaged 21-cm power spectrum given by 
\begin{equation}\label{eq:ps}
\Delta_{21}^2 = \bar{\delta T_{\rm b}}^2 \frac{k^3}{4\pi^2}\left(\langle\delta_{21}^2\rangle + \frac{2}{3}\langle\delta_{21}\delta_{m}\rangle + \frac{1}{5}\langle\delta_m^2\rangle\right),
\end{equation}
where $\delta_m$ is the matter overdensity field. We have used Eq.~(\ref{eq:dTb}) with the definition $\delta T_b = \bar{\delta T_b}(1+ \delta_{21})$, with $\bar{\delta T_b}$ being the globally averaged differential brightness temperature. Note that the first term gives the 21-cm signal in real space and the second and third terms in Eq.~(\ref{eq:ps}) are a consequence of averaging out the angular dependence introduced by the redshift space distortion term \cite{Barkana2005AFluctuations}.
See Ref.~\cite{ross2021redshift} for a detailed study of redshift space distortion at cosmic dawn.

In Fig.~\ref{fig:PS_vs_z}, we plot the 21-cm power spectra as a function of redshift at $k=0.1$ h/Mpc. The figure is structured in the same way as Fig.~\ref{fig:global_signal}; i.e., the top and bottom panels show the WCDM and FCDM scenarios, respectively, while the three columns refer to the assumptions regarding the source models (\FLOOR, \DPL, and \TRUNC). All plots show the characteristic two-peaked shape of the power spectrum at cosmic dawn, where the first peak refers to the Lyman-$\alpha$ coupling and the second peak to the heating epoch. See, e.g., Refs.~\cite{Mesinger2013,Ghara201521cmVelocities,Ross2019EvaluatingDawn} for more discussion about this.

As a general effect visible in Fig.~\ref{fig:PS_vs_z}, we see that the positions of the \lya and heating peaks move toward smaller redshifts for mixed DM models compared to CDM shown in blue. At the same time, the peaks are also shifted to smaller redshifts when photon sources are less efficient, i.e. for the case of the \TRUNC\ models compared to the \DPL\ and \FLOOR\ model. This trend again points toward potential degeneracies between DM and source models, discussed before.

Finally, it is worth noticing that the different DM models seem to be best distinguishable from each other when assuming the \FLOOR\ model (as opposed to the \DPL\ and, more so, the \TRUNC\ model). This is not surprising, since it is the small halo mass scales that contain the most information about the nature of dark matter. The more minihaloes are populated with sources, the more it will be possible to constrain noncold DM models.



\section{\label{sec:forecast} SKA forecast}

After discussing the link between the 21-cm signal and the DM sector at the level of individual benchmark models, we now perform a more detailed analysis including the full model parameter space. The basic question we address is how well the upcoming Square Kilometre Array (SKA) telescope will be able to constrain the fermionic and bosonic mixed DM scenarios. For this reason, we set up realistic mock data set from SKA-Low based on a $\Lambda$CDM cosmology and assuming instrumental noise, sample variance, and foreground contamination. With this at hand, we will perform a Markov Chain Monte Carlo (MCMC) inference analysis of the 21-cm power spectrum, simultaneously varying astrophysical and dark matter parameters. The final goal is to investigate to what extent we can go beyond current constraints for warm, fuzzy, and mixed DM models.

\subsection{Mock observations}
\label{sec:mock_obs}
We start by presenting our mock observations for SKA-Low. We fully consider the observational errors due to instrumental noise and sample variance, while the foreground contamination is accounted for by applying cuts to the mock data. This analysis is similar to previous forecast studies in the literature \cite[see e.g. Refs.][]{Greig201521CMMC:Signal,greig2017simultaneously,Park2019InferringSignal,qin2021tale}.

First, we simulate the interferometric data by considering the antennae configuration of the first phase of SKA-Low 
\footnote{The configuration used in this work is taken from \url{https://astronomers.skatelescope.org/wp-content/uploads/2016/09/SKA-TEL-SKO-0000422_02_SKA1_LowConfigurationCoordinates-1.pdf}}, which is currently being constructed in western Australia. This phase of SKA-Low is planned to be composed of $\sim 512$ radio antennae with a core area of 1 km diameter where the antennae density will be high.
The telescope configuration and observation properties, such as e.g. collecting area, declination, system temperature ($T_\mathrm{sys}(\nu)$, where $\nu$ is the observed frequency), and bandwidth, are summarized in Table~\ref{tab:telescope_config}.

\begin{table}[b]
\caption{\label{tab:telescope_config}%
The properties of the first phase of SKA-Low and observation parameter values.
}
\begin{ruledtabular}
\begin{tabular}{lc}
Parameters & Values \\
\colrule
Total number of antennas 
& 512 \\
Diameter of the core & 1 km \\
System temperature ($T_\mathrm{sys}$) & 
$\left[ 100 + 60(\frac{\nu}{300\mathrm{MHz}})^{-2.55}\right]$ K \\
Effective collecting area 
& 962 $\mathrm{m}^2$ \\
Declination & -30$^\circ$ \\
Total observation time & 1000 h \\
Observation hour per day & 6 h \\
Signal integration time & 10 s \\
Bandwidth & 10 MHz \\
\end{tabular}
\end{ruledtabular}
\end{table}

 \begin{figure*}[t] 
 \centering
  \includegraphics[width=0.95\textwidth]{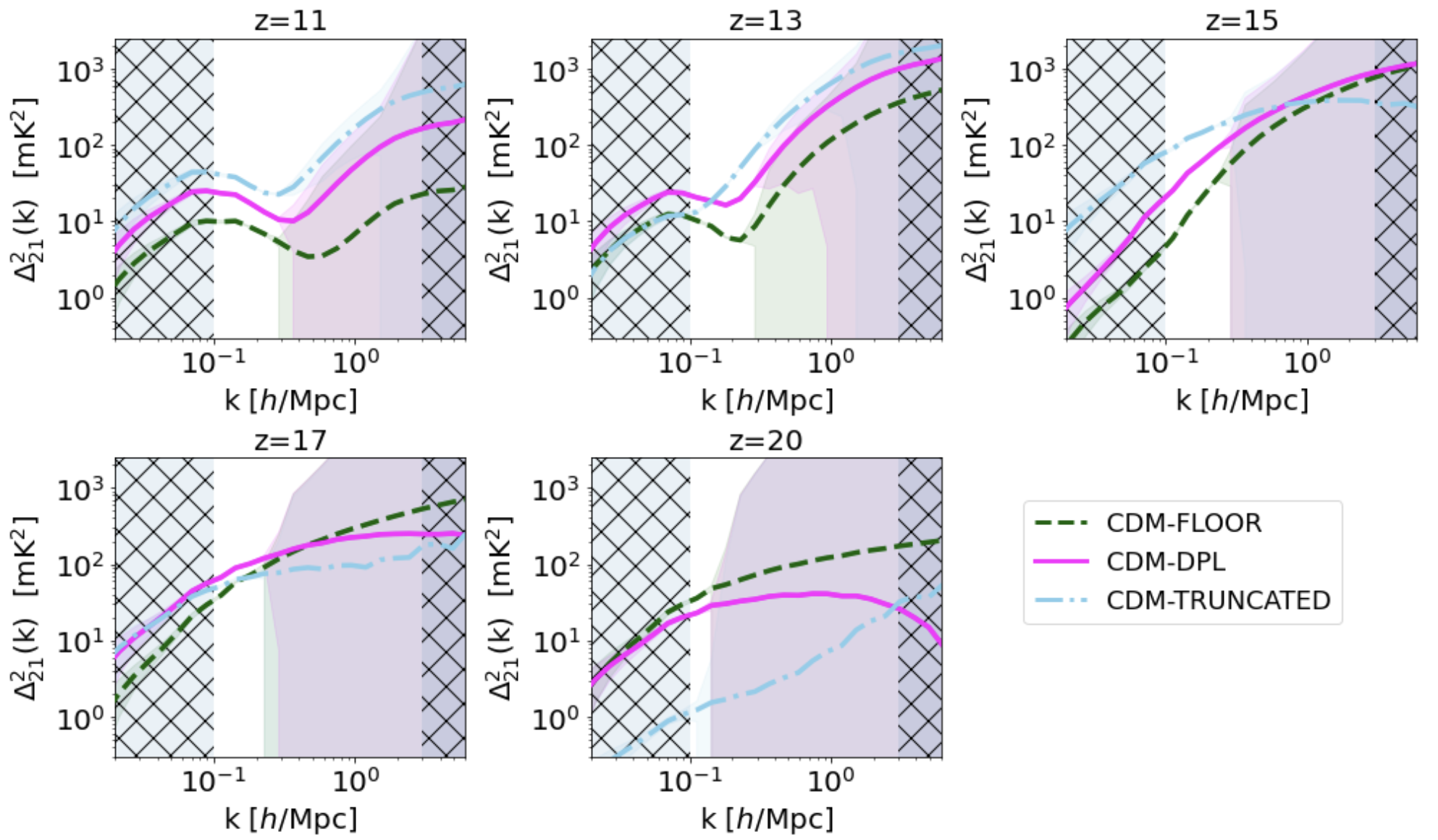}
 \caption{The 21-cm power spectrum for CDM universe at different redshifts assuming the source models and error components used in the mock data set. The purple, orange, and green lines correspond to the \FLOOR, \DPL, and \TRUNC\ source models, respectively. The color-shaded areas show the errors due to instrumental noise and sample variance. The hatched areas indicate scale cuts due to foreground contamination and noise. See text for  more information.
 }
 	\label{fig:mock_obs_multiz}
 \end{figure*}
 
 

For a symmetric radial distribution of antennae, the error on the power spectrum due to instrumental noise can be calculated analytically \cite[see e.g.,][]{Mellema2013ReionizationArray,Koopmans2015TheArray}. However, the true antennae distribution is nonuniform. An interferometry-based radio telescope records the complex visibilities, which are proportional to the Fourier transform of the signal in the real space. These visibilities are observed in the \textit{uv} space. The telescope samples the \textit{uv} space due to the rotation of Earth. See, e.g., Ref.~\cite{Rohlfs2013ToolsAstronomy} for detailed description of interferometric observations. 

We use the \texttt{Tools21cm} code \cite{Giri2020tools21cm} to track the \textit{uv}-space sampling (or \textit{uv} coverage) assuming a daily observation of 6 hours (see, e.g., Fig.~2 in Ref.~\cite{prelogovic2021machine} for the simulated \textit{uv} coverage of the first phase of SKA-Low). We assume an integration time of 10 s, which is the time after which the observation is recorded. For a detailed description of the method, see Refs.~\cite{Ghara2017ImagingSKA,giri2018optimal}.

\texttt{Tools21cm} outputs realizations of the instrumental noise for a given interferometric telescope by taking into account the daily \textit{uv} coverage. The instrumental noise level is inversely related to the bandwidth of the observations and the total observation time. We simulate observations assuming 1000 h exposure with SKA-Low. Previous studies have shown that with SKA-Low, such an observation time will yield a sufficient signal-to-noise ratio (SNR) to study the properties of cosmic dawn \cite[e.g.][]{Mellema2013ReionizationArray,Koopmans2015TheArray,Mellema2015HISKA,greig2017simultaneously,Ross2019EvaluatingDawn,ross2021redshift}. We furthermore assume a bandwidth of 10 MHz. Within this bandwidth, the data will not be significantly affected by the light-cone effect studied in Refs.~\cite{Datta2014LightImplications,Ghara201521Effects,Giri2018BubbleTomography}. The error from instrumental noise is calculated using the procedure described in Sec. 3.2.1 of  Ref.~\cite{Jensen2013ProbingDistortions}. As part of this procedure, we numerically simulate many realizations ($\sim 100$) of the instrumental noise to estimate the standard deviation ($\sigma_{21}$) on the power spectrum at various wave modes.


The error due to the sample variance is included using Eq.~(9) in Ref.~\cite{Mellema2013ReionizationArray}. It depends on the field-of-view and therefore on the beam width of the telescope. The size of the beam is given by the diameter of the antenna stations, which is $\sim 35$ m for the first phase of SKA-Low.

The foreground contamination is not modeled in detail here. Instead, we assume the foreground signal to reside in the foreground wedge \cite[see e.g.][]{Liu2014EpochFormalism,Barry2016CalibrationSKA}. Following Refs.~\cite{pober2014next,Greig201521CMMC:Signal}, we apply a general, redshift independent cut at $k=0.1$ h/Mpc. 
We also discard data from above $k=3$ h/Mpc. These scales are expected to be unusable due to very large instrumental noise with our instrumental setup \cite[e.g.][]{ross2021redshift}. 
For our mock observations, we assume a $\Lambda$CDM cosmology. This allows us to quantify the constraining power of the 21-cm signal with respect to mixed DM model parameters. 

Another important ingredient of our mock data is the source modeling. In order to account for current uncertainties, especially at the low-mass range, we include mocks for all three source models (\FLOOR, \DPL\ and \TRUNC) shown in Fig.~\ref{fig:source_models}. The \FLOOR\ model corresponds to an optimistic case in which the small-mass sources are highly efficient, whereas the \TRUNC\ model shows a pessimistic case with no sources below the atomic cooling limit ($\sim 10^8 M_\odot$). The \DPL\ model lies in between, allowing for sources below the atomic cooling limit, but assuming very inefficient star-formation rates.

The mock data are set up assuming five different redshift bins between $z=11$ and 20, 
each of them with 15 data points equally separated in log space within the range $k= 0.1-3.0~$h/Mpc. These bins are selected to be broad enough to minimize correlations between data points in redshift and $k$ space. This considerably simplifies our analysis, as we do not have to deal with nondiagonal terms in the covariance matrix during the inference analysis.


In Fig.~\ref{fig:mock_obs_multiz}, we show the 21-cm mock power spectra for the five redshift bins and assuming the three different source models: \FLOOR\ (dashed dark green lines), \DPL\ (solid magenta lines), and \TRUNC\ (dotted-dashed sky blue lines). The error due to instrumental noise and sample variance are indicated by the color-shaded areas, which have been added in quadrature. While the former induces a strong increase toward high wave numbers, the latter leads to a gradual increase toward large scales, mainly visible at $k<0.1$ h/Mpc. The hard scale cuts due to foreground contamination and noise are shown as hatched gray areas.

 \begin{figure*}[t] 
 \centering
 \begin{tabular}{c c}
 ~~~~~~~~~~~~~~~\textbf{\Large WCDM} & ~~~~~~~~~~~~~~~\textbf{\Large FCDM} \\
  \includegraphics[width=0.43\textwidth]{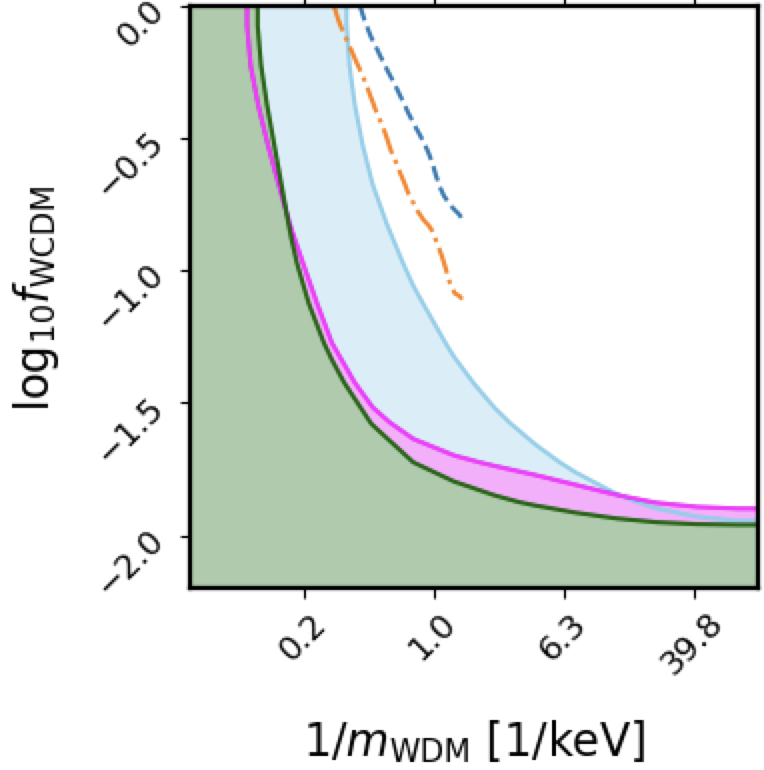} &
  \includegraphics[width=0.43\textwidth]{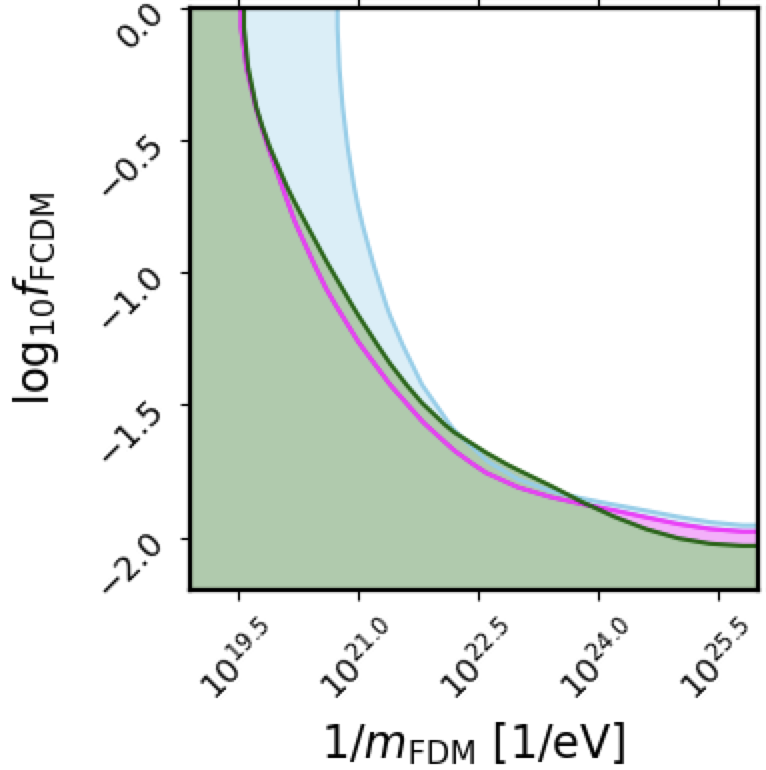} 
  \end{tabular} \\
  \includegraphics[width=0.2\textwidth]{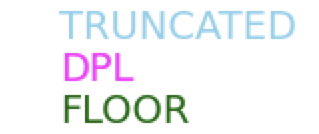} 
  \includegraphics[width=0.3\textwidth]{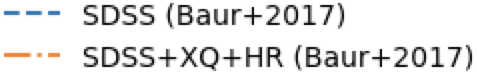} \\
 \caption{Forecast constraints on the particle mass and fraction of a fermionic (left) and bosonic (right) mixed DM scenario. Different colors correspond to different assumptions regarding the underlying source model of the mock data set (\TRUNC\ in sky blue, \DPL\ in magenta, and \FLOOR\ in dark green). All constraints are provided at the 95\% confidence level.
 }
 	\label{fig:cornerplot_CWDMmodel_CFDMmodel_CDMobs}
 \end{figure*}
 
Accounting for all uncertainties, Fig.~\ref{fig:mock_obs_multiz} shows that we can expect to observe the 21-cm power spectrum up to $z\approx 20$ with 1000 h data from the first phase of SKA-Low. The most interesting window, showing minimal noise contamination, lies between $k\sim 0.1$ h/Mpc and $k\sim 0.3-1$ h/Mpc and becomes narrower with increasing redshift. Note, furthermore, that at a given redshift and $k$ value, the detectability of the signal will ultimately depend on its strength, which is a function of cosmology and the underlying astrophysics.



\subsection{Inference analysis and constraints}
\label{sec:mock_obs_results}




\begin{table}[b]
\caption{\label{tab:mockobs_model}%
The prior range explored in the forecast study. We have four parameters ($f_{\star,0}, \gamma_4, f_X, E_0$) to describe the photon sources. Our mixed DM models are given by two parameters ($m_{n\mathrm{DM}}, f_{n\mathrm{CDM}}$).
}
\begin{ruledtabular}
\begin{tabular}{lcccc}
Parameters & Prior range & \FLOOR\ & \DPL\ & \TRUNC\ \\
\colrule 
$f_{\star,0}$ & $[0.1, 0.5]$ & 0.3 & 0.3 & 0.3 \\
$\gamma_4$ & $[-5, 5]$ & 0.2 & 0 & -3 \\
$f_{X}$ & $[0.05, 0.5]$ & 0.2 & 0.2 & 0.2 \\
$E_0$/eV & $[100, 1000]$ & 500 & 500 & 500\\
\hline
log$_\mathrm{10}$(keV$/m_\mathrm{WDM}$) & $[-3, 2]$ & -3 & -3 & -3 \\
log$_\mathrm{10}$($f_\mathrm{WDM}$) & $[-3, 0]$ & -3 & -3 & -3 \\
\hline
log$_\mathrm{10}(\mathrm{eV}/m_\mathrm{FDM}$) & $[18, 26]$ & 18 & 18 & 18 \\
log$_\mathrm{10}f_\mathrm{FDM}$ & $[-3, 0]$ & -3 & -3 & -3 \\
\end{tabular}
\end{ruledtabular}
\end{table}

 

 

As mentioned above, we use a MCMC framework to determine the constraints on our model parameters $\pmb{\theta}$. The posterior distribution of $\pmb{\theta}$, given the observational data $\pmb{x}$, is
\begin{eqnarray}
 p(\pmb{\theta}|\pmb{x}) \propto \mathcal{L}(\pmb{\theta}) \pi(\pmb{\theta}) \ ,
\end{eqnarray}
where $\pi(\pmb{\theta})$ is the prior probability of the model parameters, and $\mathcal{L}(\pmb{\theta}) = p(\pmb{x}|\pmb{\theta})$ is the likelihood function. We use the log likelihood given by
\begin{eqnarray}
\log \mathcal{L} = - \frac{1}{2} \sum_{k,z} \left( \frac{\Delta^2_{\rm 21,ob}(k,z)-\Delta^2_{\rm 21,th}(k,z)}{\sigma_{21}(k,z)^2} \right)^2,
\label{eq:likelihood}
\end{eqnarray}
where $\Delta^2_{\rm 21,ob}$ and $\Delta^2_{\rm 21,th}$ are the observed (mock) and modeled power spectra, respectively, and $\sigma_{21}$ is the combined error on the data. 

In terms of the modeling, we vary the six parameters $\pmb{\theta}=\lbrace f_{\rm nCDM},\,m_{\rm nCDM}, \,f_{\star,0},\,\gamma_4, \,f_X, \,E_0\rbrace$. Next to the two dark matter parameters ($f_{\rm nCDM}$, $m_{\rm nCDM}$), we include four source parameters in our MCMC analyses. They are related to the overall amplitude ($f_{\star,0}$) and the small-scale behaviour ($\gamma_4$) of the stellar-to-halo mass relation as well as the amplitude ($f_X$) and the spectral range ($E_0$) of the x-ray flux. Other astrophysical parameters are kept fixed for simplicity.
\RefReport{Previous works have shown that the signal is sensitive to the atomic cooling threshold \cite[e.g.][]{greig2017simultaneously}. As we study haloes below the atomic cooling limit, $\gamma_4$ mimics the effect by suppressing star formation in small mass haloes.}

The prior ranges along with the mock values (for the three source models) are provided in Table~\ref{tab:mockobs_model}. Note that we assume  uniform priors for the four astrophysical parameters and uniform priors in log space for the two dark matter parameters.

In Fig.~\ref{fig:cornerplot_CWDMmodel_CFDMmodel_CDMobs}, we plot the constraints on the WCDM (left panel) and the FCDM (right panel) model parameters at 95 percent confidence level based on the three different source models for the mock data (presented with different colors). Note that all astrophysical parameters are marginalized over. Full corner plots including posterior contours of all astrophysical parameters for both WCDM and FCDM cases are provided in Appendix~\ref{appx:mcmc_full}.

\begin{table}[t]
\caption{\label{tab:mcmc_summary} Forecast constraints from special locations of the mixed DM parameter space: pure warm DM, pure fuzzy DM, CDM plus a subdominant fermionic relic, CDM plus a subdominant bosonic relic. All constraints are provided for the \FLOOR, the \DPL\, and the \TRUNC\ source models. They are given at the 95\% confidence level. 
}
\begin{ruledtabular}
\begin{tabular}{lr}
Source model: \FLOOR\ & \\
\hline
Warm DM ($f_{\rm WCDM}\sim 1$) & $m_{\rm WDM}\gtrsim 15$ keV \\
Fuzzy DM ($f_{\rm FCDM}\sim 1$) & $m_{\rm FDM}\gtrsim 2\times 10^{-20}$ eV \\
CDM + hot fermionic relic & $f_{\rm WCDM}\lesssim0.012$\\
CDM + hot bosonic relic & $f_{\rm FCDM}\lesssim 0.010$\\
\hline\hline
Source model: \DPL\ & \\
\hline
Warm DM ($f_{\rm WCDM}\sim 1$)  & $m_{\rm WDM}\gtrsim 15$ keV \\
Fuzzy DM ($f_{\rm FCDM}\sim 1$)  & $m_{\rm FDM}\gtrsim 2\times10^{-20}$ eV \\
CDM + hot fermionic relic & $f_{\rm WCDM}\lesssim 0.012$\\
CDM + hot bosonic relic & $f_{\rm FCDM}\lesssim 0.011$\\
\hline\hline
Source model: \TRUNC\ & \\
\hline
Warm DM ($f_{\rm WCDM}\sim 1$) & $m_{\rm WDM}\gtrsim 4$ keV \\
Fuzzy DM ($f_{\rm FCDM}\sim 1$) & $m_{\rm WDM}\gtrsim 2\times10^{-21}$ eV \\
CDM + hot fermionic relic & $f_{\rm WCDM}\lesssim 0.012$\\
CDM + hot bosonic relic & $f_{\rm FCDM}\lesssim 0.011$\\
\end{tabular}
\end{ruledtabular}
\end{table}

The DM constraints behave as expected in the sense that the parameter space with large $f_{\rm nCDM}$ and small $m_{\rm nCDM}$ in the top-right corner of the plots is strongly ruled out. Very small values of $f_{\rm nCDM}$ (bottom part of the plots), on the other hand, remain allowed independently of $m_{\rm nCDM}$. The same is true for large values of $m_{\rm nCDM}$ (left-hand side of the plots) that stay allowed independently of the fraction $f_{\rm nCDM}$.

Focusing on the different source models, constraints are weakest for the \TRUNC\ model (sky blue contours). This is expected as the model lacks sources in small haloes, where the mixed DM models differ the most from $\Lambda$CDM. The \DPL\ and \FLOOR\ models (magenta and dark green contours) yield considerably stronger constraints, especially in the regime of large $f_{\rm nCDM}$. Interestingly, they hardly differ between each other in their constraining power. This points toward the possibility that minihaloes, as long as they are populated with sources, are capable of providing stringent constraints on DM parameters. It does not seem to matter much how large their star-formation rate effectively is.

It is also worth noticing that in the regime of small $m_{\rm nCDM}$ (bottom right part of the plots) for both WCDM and FCDM models, the constraints are approximately independent of the source models (with all contours lying on top of each other). This is due to the fact that models with small $f_{\rm nCDM}$ and $m_{\rm nCDM}$ exhibit halo mass functions that are modified over all mass scales and are therefore not solely dependent on the abundance of sources in minihaloes. 

\RefReport{In the left panel of Fig.~\ref{fig:cornerplot_CWDMmodel_CFDMmodel_CDMobs}, we show the constraints given in Ref.~\cite{baur2017constraints} from Lyman-$\alpha$ forest observations. 
The dashed and dash-dotted lines show the constraints from SDSS data alone and from a combination of SDSS data with XShooter, MIKE and HIRES data. This latter data set includes much smaller scales and is, therefore, more constraining. See Refs.~\cite{baur2017constraints,garzilli2019warm} for a discussion on the validity of these results.}

Some of the constraints provided in Fig.~\ref{fig:cornerplot_CWDMmodel_CFDMmodel_CDMobs} are considerably stronger than current constraints from the literature. In the limit of pure WDM and FDM ($f_{\rm nCDM}\sim 1$), for example, we obtain constraints of $m_{\rm WDM}\gtrsim 4,\,15,\,15$ keV and $m_{\rm FDM}\gtrsim 2\times10^{-21},\,2\times10^{-20},\,2\times10^{-20}$ eV for the \TRUNC, \DPL, and \FLOOR\ model, respectively. In the limit of very small particle masses, on the other hand, we obtain limits on the fraction of $f_{\rm WCDM}\lesssim 0.01$ and $f_{\rm FCDM}\lesssim 0.01$ which do not depend on the source model. 
All these constraints are at 95\% confidence level. 
A summary of these constraints are provided in Table ~\ref{tab:mcmc_summary}.

Some previous studies have included a theoretical error in their inference pipeline to account for the limitations of current modeling methods \cite[e.g.][]{Greig201521CMMC:Signal,Ghara2020ConstrainingObservations,Greig2021ExploringSignal,Ghara2021ConstrainingObservations}. In this paper, however, we neglect all modeling uncertainties despite the fact that they are currently larger than the expected errors from SKA-Low. Note that this is common practice for forecast studies. It means that, until the SKA data are available, we expect the current prediction techniques to improve to the level that the theoretical errors become subdominant. In Appendix~\ref{appx:mcmc_model_error}, we nevertheless show the impact of a theoretical error on the DM parameter constraints.

Here, we have quantified the constraining power of SKA-Low on mixed DM scenarios provided the true Universe is filled with pure CDM. We have shown that such as setup will lead to unprecedented constraints on the particle mass and fraction of fermionic and bosonic noncold relics, especially for the case that minihaloes are populated with stars. However, we cannot say anything about constraints if the true model is not CDM but mixed DM instead. In that case, we expect the observable signal from SKA-Low to be pushed to smaller redshifts. Such a signal will be more difficult to interpret due to the strong degeneracies between DM and source model parameters.


\section{\label{sec:summary} Summary}
 
In this paper, we have extended the framework from Ref.~\cite{Schneider2021HaloDawn} to explore nonstandard DM models using the 21-cm signal at cosmic dawn. We have focused specifically on fermionic and bosonic mixed DM scenarios characterized by the noncold particle mass $m_{\rm nCDM}$ and the mixing fraction $f_{\rm nCDM}$. In the case of fermionic mixed DM, the noncold component could be made of a sterile neutrino with mass in the eV or keV range. In the bosonic case, the noncold component consists of a ultralight axion DM particle with mass in the range of $\sim 10^{-18}-10^{-26}$ eV.

In terms of structure formation, mixed DM scenarios lead to modifications of the matter power spectrum, the halo mass function, and the mass accretion rates. This affects both the clustering of the gas as well as the properties of the sources. As a result, the characteristic double-peak feature of the 21-cm power spectrum at cosmic dawn is pushed to smaller redshifts and exhibits an increased amplitude compared to the standard $\Lambda$CDM model.

The main task of the paper consists of performing a forecast analysis based on the low-frequency array of the upcoming Square Kilometre Array telescope (SKA-Low). We carried out an inference analysis based on the Markov Chain Monte Carlo (MCMC) method, varying two DM and four astrophysical parameters. For the mock data set, we investigated three choices for the source modeling, a pessimistic, neutral, and optimistic one. They differ with respect to the assumed abundance of sources in minihaloes, going from zero to efficient star formation below the atomic cooling limit.

For the SKA forecast analysis, we assume a $\Lambda$CDM universe, investigating the potential of constraining mixed DM scenarios. In the case that minihaloes are populated with stars, we obtained stringent constraints that go far beyond current limits. If minihaloes are devoid of stars, the constraints become weaker but remain more stringent than the current limits from the Lyman-$\alpha$ data, Milky Way satellite counts, or strong lensing studies. A summary of the constraints are provided in Fig.~\ref{fig:cornerplot_CWDMmodel_CFDMmodel_CDMobs} and Table \ref{tab:mcmc_summary}.

Interestingly, the DM constraints do not increase linearly with the number of stars in minihaloes. We rather find that the constraints remain strong as long as there is at least some star formation in minihaloes, even if the stellar-to-halo ratio is strongly suppressed. This finding is encouraging because we expect at least some stars to form in haloes between $10^6-10^8$ $M_{\odot}$ via the molecular cooling channel, even when assuming efficient Lyman-Werner feedback process.

In summary, the results from this paper indicate that the SKA telescope has the potential to significantly push current constraints on warm, fuzzy, and mixed DM scenarios. However, the true strength of these constraints will ultimately depend on whether the SKA will provide evidence for star-formation in minihaloes below the atomic cooling limit. If this is not the case, then it will be difficult to distinguish between an inefficient source models and a Universe filled with warm, fuzzy, or mixed dark matter.

\begin{acknowledgments}
This work is
supported by the Swiss National Science Foundation via
the Grant No. {\sc PCEFP2\_181157}.
All the MCMC runs were done using the {\tt emcee} \cite{foreman2013emcee} package.
For analysis and plotting, we have used the following software packages: {\tt numpy} \cite{van2011numpy}, {\tt scipy} \cite{virtanen2020scipy}, {\tt matplotlib} \cite{hunter2007matplotlib} and {\tt corner} \cite{corner}.
\end{acknowledgments}

\appendix

\section{\label{appx:mcmc_full} COMPLETE POSTERIOR DISTRIBUTION FROM THE MCMC RUNS}

 
 \begin{figure*}[t] 
 \centering
 \textbf{\Large WCDM}\\
  \includegraphics[width=0.71\textwidth]{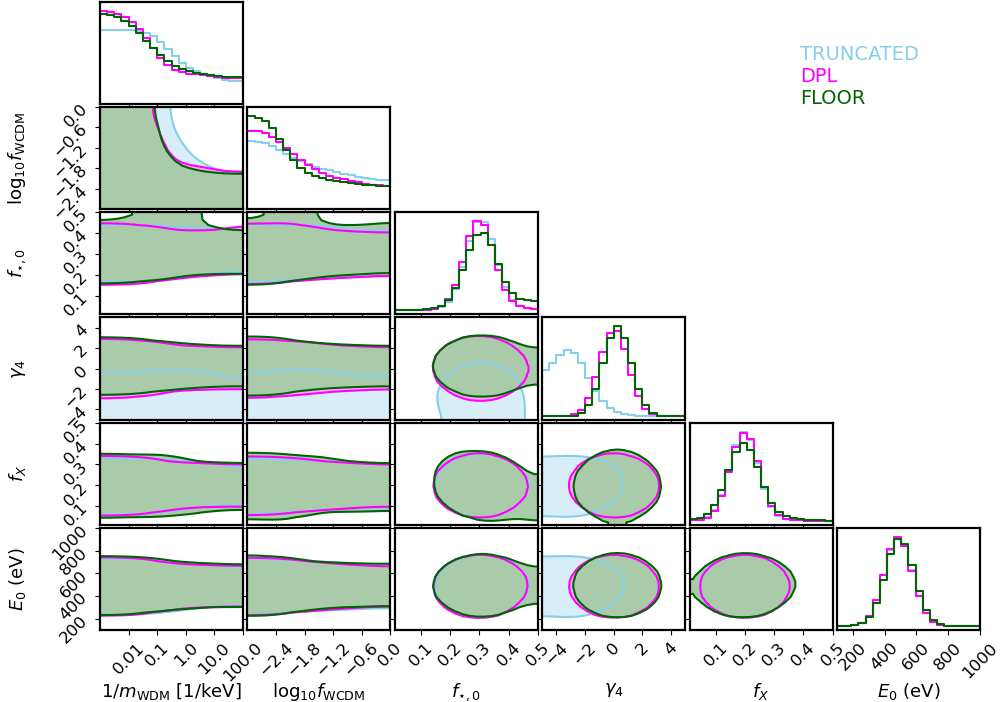}\\ \vspace{0.3cm}  
  \textbf{\Large FCDM}\\
  \includegraphics[width=0.71\textwidth]{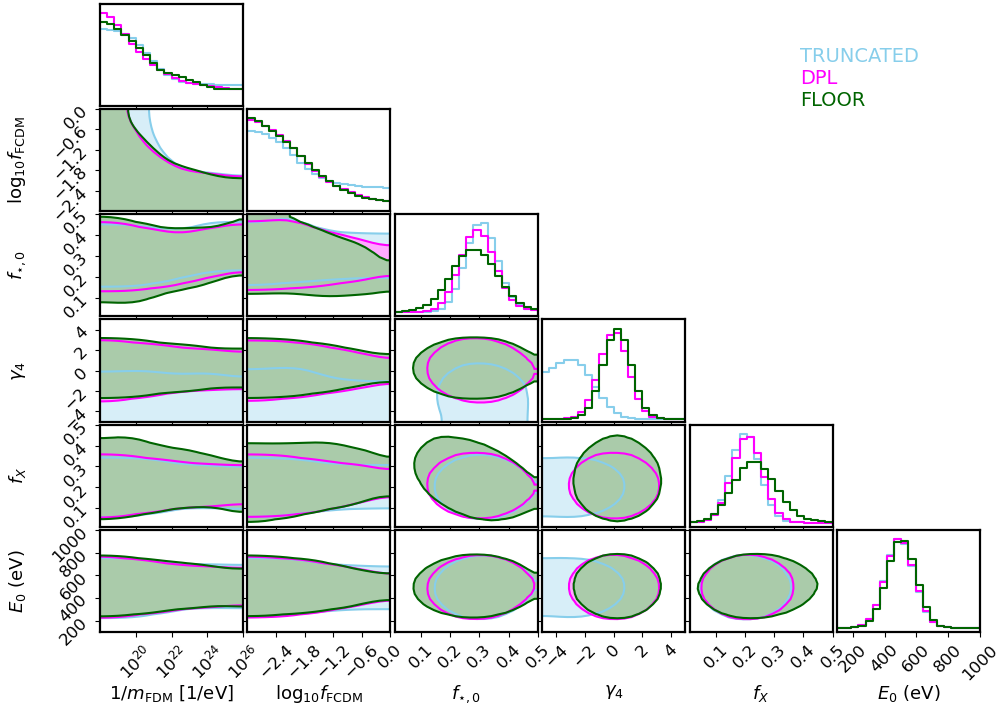}
 \caption{Corner plots showing the 95\% confidence interval of the posterior distributions on six parameter WCDM (top) and FCDM (bottom) models. As in the main text, we assume the source models \TRUNC\ (sky blue), \DPL\ (magenta) and \FLOOR\ (dark green) for the mock data. The peaks of the 1D marginalized probability distributions agree with the true values of the mock observations for all cases.
 }
 \label{fig:cornerplot_CFDMmodel_CDMobs_full}
 \end{figure*}

The primary aim of this work is to study the capability of SKA in constraining mixed dark matter (DM) models. Therefore, we only present the posterior distributions of the DM parameters in the main text. However, while analyzing the 21-cm signal observations, we can also constrain the astrophysical processes driving the Lyman-$\alpha$ coupling and heating of the hydrogen gas in the IGM during cosmic dawn.

In our MCMC analyses (Sec.~\ref{sec:mock_obs_results}), we vary four astrophysical parameters ($f_{\star,0}, \gamma_4, f_X, E_0$)  along with the two DM parameters ($m_{\rm nCDM}, f_{\rm nCDM}$). For completeness, we show the posterior distributions on all parameters in Fig.~\ref{fig:cornerplot_CFDMmodel_CDMobs_full}.  Note that all contours are shown at the 95\% confidence level.

With all three mock observations, which are the \TRUNC\ (sky blue contour), \DPL\ (magenta contour), and \FLOOR\ (dark green contour), we are able to constrain the astrophysical parameters quite well. The peak of the 1D marginalized probability distribution function of all the astrophysical parameters lie very close to the mock values for both WCDM and FCDM scenarios. This is an important cross-check validating the outcome of our inference analysis.

The results of Fig.~\ref{fig:cornerplot_CFDMmodel_CDMobs_full} show furthermore that with data from SKA-Low, it will be possible to gain much better knowledge of the underlying astrophysical model. In particular, it will be possible to derive strong constraints on the stellar-to-halo mass relation and the x-ray efficiency during the epoch of cosmic dawn.

\section{\label{appx:mcmc_model_error} IMPACT OF MODELING ERROR ON THE CONSTRAINTS}


 \begin{figure*}[t] 
 \centering
  \includegraphics[width=0.45\textwidth]{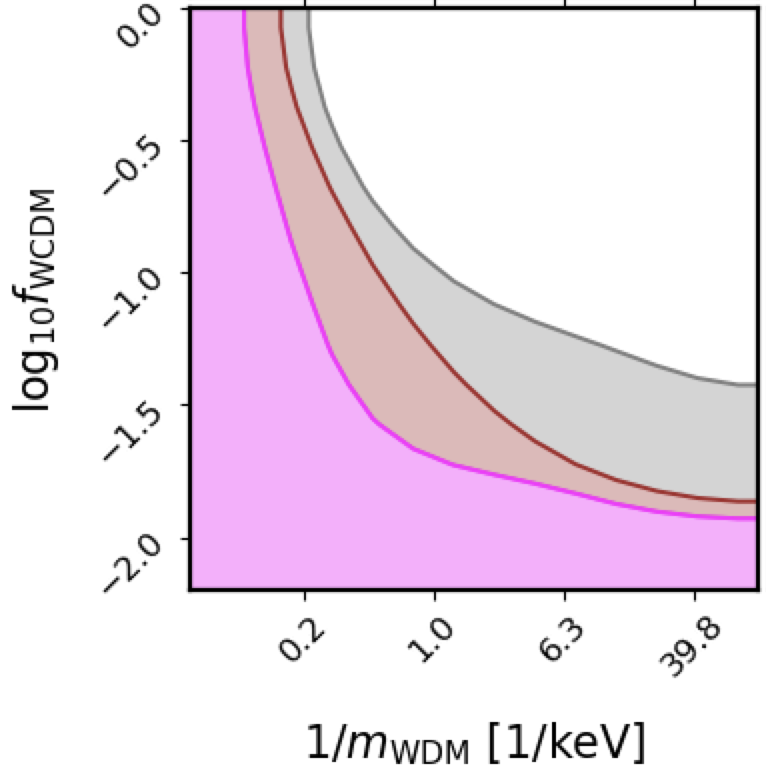}
  \includegraphics[width=0.20\textwidth]{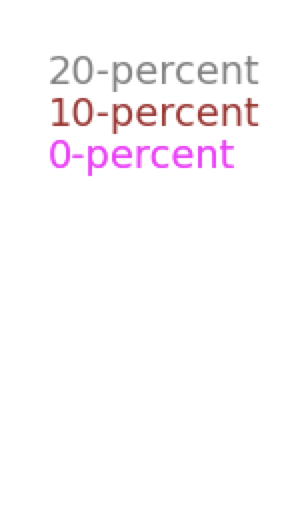}
 \caption{Forecast constraints on the WCDM model parameters assuming a theoretical modeling error of 0\%, 10\% and 20\% (magenta, brown and gray). Here we assumed the \DPL\ source model. All constraints are provided at the 95\% confidence level. 
 }
 	\label{fig:cornerplot_MDMmodel_DPL_moderrs_CDMobs}
 \end{figure*}
 
In many studies analyzing the 21-cm power spectrum, an additional modeling error is introduced to account for the current limitations of signal simulation methods \cite[e.g.][]{Greig201521CMMC:Signal,Ghara2020ConstrainingObservations,Greig2021ExploringSignal,Ghara2021ConstrainingObservations}. Since our work is limited to a forecast analysis, we do not include any modeling errors in the analysis presented in the main text. This implicitly means that we expect current prediction techniques to improve until the observations from the SKA telescope will be available.

In this appendix, we study the impact of modeling errors for the case that the improvement of prediction techniques does not proceed as expected. Similar to previous methods from the literature, we empirically define the $m$ percent modeling error as the standard deviation of $m$ percent at every wave-mode and redshift. This error is appended in quadrature to the variance ($\sigma_{21}$) shown in Eq.~(\ref{eq:likelihood}).

In Fig.~\ref{fig:cornerplot_MDMmodel_DPL_moderrs_CDMobs}, we show the constraints on the WCDM models, assuming a modeling error of 0\% (magenta), 10\% (brown), and 20\% (gray). As mock data set, we consider the \DPL\ source model. When the DM is purely noncold ($f_{\rm WCDM}\sim 1$), the particle mass is constrained to be greater than 15, 9 and 6 keV for 0\%, 10\% and 20\% modeling error respectively. In the other extreme case where the particle mass is very small, we constrain the fraction of noncold DM to be less than 0.012, 0.014 and 0.038 for 0\%, 10\% and 20\% modeling error respectively. These constraints are given at 95\% confidence level.

We see that the constraints become weaker as we increase the modeling error. However, even with 20\% modeling error, we get competitive constraints on the particle mass and fraction of mixed dark matter. Here we empirically assumed the modeling error. In the future, we will compare our simulation framework with other methods to better understand this error.



\bibliography{refs,mendeley}

\end{document}